\documentclass[aps,prl,reprint,groupedaddress,superscriptaddress]{revtex4-2}

\usepackage[utf8]{inputenc}
\usepackage[T1]{fontenc}
\usepackage[german,english]{babel}

\usepackage{amsmath}
\usepackage{mathtools}
\usepackage{amssymb}
\usepackage{amsthm}
\usepackage{microtype} 
\usepackage{quoting} 
\quotingsetup{font=small}

\usepackage[pdftex]{graphicx}
\usepackage{fancyhdr}
\usepackage{bm}
\usepackage{comment}
\usepackage{booktabs}
\usepackage{eurosym} 
\usepackage{braket}
\usepackage{bbold} 
\usepackage{emptypage}
\usepackage{epstopdf}
\usepackage{verbatim}
\usepackage{enumitem}
\usepackage{float} 
\usepackage{chemformula}

\usepackage{commath}

\usepackage{color}

\definecolor{Blue}{rgb}{0.00, 0.00, 1.00}
\definecolor{Red}{rgb}{1.00, 0.00, 0.00}

\newcommand{\gsim}{\raisebox{-0.13cm}{~\shortstack{$>$ \\[-0.07cm]
      $\sim$}}~}
\newcommand{\RomanNumeralCaps}[1]
    {\MakeUppercase{\romannumeral #1}}

\DeclareMathOperator{\Tr}{Tr}
\DeclareUnicodeCharacter{2212}{-}
\DeclareMathOperator\erf{erf}

\usepackage[bookmarks=true,colorlinks,linkcolor=blue,urlcolor=blue,citecolor=blue]{hyperref} 

\begin{document}

\title{Emergent quantum correlations and collective behavior  in non-interacting quantum systems subject to stochastic resetting}

\author{Matteo Magoni}
\affiliation{Institut für Theoretische Physik, Eberhard Karls Universität Tübingen, Auf der Morgenstelle 14, 72076 Tübingen, Germany}

\author{Federico Carollo}
\affiliation{Institut für Theoretische Physik, Eberhard Karls Universität Tübingen, Auf der Morgenstelle 14, 72076 Tübingen, Germany}

\author{Gabriele Perfetto}
\affiliation{Institut für Theoretische Physik, Eberhard Karls Universität Tübingen, Auf der Morgenstelle 14, 72076 Tübingen, Germany}

\author{Igor Lesanovsky}
\affiliation{Institut für Theoretische Physik, Eberhard Karls Universität Tübingen, Auf der Morgenstelle 14, 72076 Tübingen, Germany}
\affiliation{School of Physics and Astronomy and Centre for the Mathematics and Theoretical Physics of Quantum Non-Equilibrium Systems, The University of Nottingham, Nottingham, NG7 2RD, United Kingdom}

\date{\today}

\begin{abstract}
We investigate the dynamics of a non-interacting spin system, undergoing coherent Rabi oscillations, in the presence of stochastic resetting. We show that resetting generally induces long-range quantum and classical correlations both in the emergent dissipative dynamics and in the non-equilibrium stationary state. Moreover, for the case of conditional reset protocols --- where the system is reinitialized to a state dependent on the outcome of a preceding measurement --- we show that, in the thermodynamic limit, the spin system can feature collective behavior which results in a  phenomenology reminiscent of that occurring in non-equilibrium phase transitions.   The discussed reset protocols can be implemented on quantum simulators and quantum devices that permit fast measurement and readout of macroscopic observables, such as the magnetisation. Our approach does not require the control of coherent interactions and may therefore highlight a route towards a simple and robust creation of quantum correlations and collective non-equilibrium states, with potential applications in quantum enhanced metrology and sensing.
\end{abstract}






\maketitle

\textbf{Introduction.---} 
Understanding and exploiting the interplay between coherent unitary evolution and measurement in quantum systems has been a central topic since the early days of quantum mechanics \cite{Born1926,BOHR1928}. Recent research in this direction is closely linked to the physics of open quantum systems \cite{Lindblad_1976,gorini_1976,gardiner_zoller,petruccione}, where interactions among quantum particles compete with the coupling to the surrounding environment. Modern experiments allow to externally control and even artificially engineer open system dynamics. This can, e.g., be achieved through so-called feedback protocols \cite{wiseman_feedback,wiseman_milburn_2009,jacobs_2014,lammers_2016,nurdin_yamamoto_2017,Kroeger_2020}, which rely on the continuous monitoring of a system followed by some action conditioned on the output of a detector. This procedure can generate non-equilibrium steady states (NESS) that feature non-trivial quantum correlations \cite{ivanov_2020,Buonaiuto_2021,ivanov_2021,young_2021}. Another approach that relies on externally imposed interventions in order to create effectively open system dynamics is \textit{stochastic resetting} \cite{stoc_reset_review}. In its simplest form it amounts to resetting a system to its initial state at random times. This procedure has been originally studied for classical diffusive systems \cite{diffusion_reset,optimal_reset,evans2014diffusion,DPTreset2015}, search processes \cite{diffusion_reset,optimal_reset,Kusmierz_2014,Pal_2017,Chechkin_2018,radice2022diffusion} and active systems \cite{evans2013optimal,christou2015diffusion,Slowman_2016,Evans_2018,Kumar_2020,Santra_2020,Bressloff_2020}, and also here interesting NESS have been shown to emerge \cite{Gupta_2014,Eule_2016,mendez_2016,Grange_2020,magoni_ising_reset,aron_2020,SokolovErgodicity_reset,Stojkoski_2021,Santra_2021,Huang_2021,Goswami_2021,Chelminiak_2021}. Similar observations have been made recently in the context of quantum systems \cite{Hartmann2006,Linden2010,Armin2020,Mukherjee_2018,Rose_2018,carollo2019unravelling,Gabriele_reset,riera_2020,Turkeshi_2021,thermodynamics_reset}. However, it remains an open question whether resetting can induce non-trivial NESS, that may display emergent quantum correlations or even non-equilibrium phase transition behavior.

\begin{figure}
    \centering
    \includegraphics[width=\linewidth]{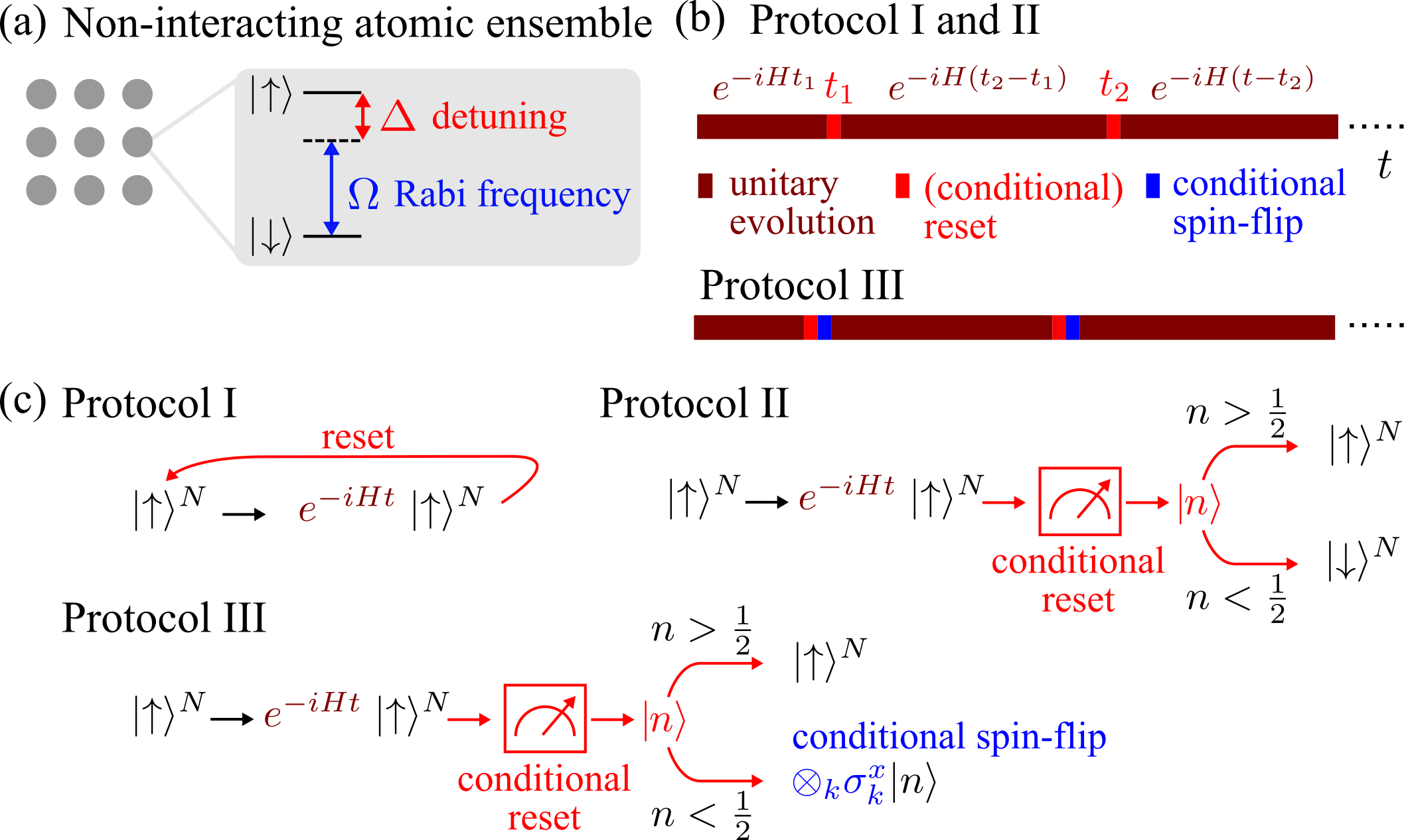}
    \caption{\textbf{Non-interacting spins subject to resetting.} (a) Non-interacting spin system subject to a (laser) field with Rabi frequency $\Omega$ and detuning $\Delta$. (b) The unitary time evolution according to Hamiltonian~\eqref{mt:Ham} is interspersed by randomly distributed reset events, which reinitialize the system to a specific state depending on the adopted reset protocol. In the figure, $t$ denotes the observation time and $t_k$ the time when the $k^\mathrm{th}$ reset event takes place. (c) Details of the reset protocols. In Protocol \RomanNumeralCaps{1}, the system is unconditionally reset to the state $\ket{\uparrow}^N$. In Protocols \RomanNumeralCaps{2} and \RomanNumeralCaps{3}, the reset is preceded by a measurement of the excitation density $n$. In Protocol \RomanNumeralCaps{2}, the value of $n$ determines the choice between two fixed reset states. In Protocol \RomanNumeralCaps{3}, when $n<1/2$, the reset state is determined by a spin flip operation applied to the state obtained from the projective measurement.}
    \label{fig:fig1}
\end{figure}

In this manuscript, we fill this gap by investigating the interplay between stochastic resetting and many-body quantum coherent evolution in the simplest --- yet surprisingly non-trivial --- case of non-interacting spin systems, see Fig.~\ref{fig:fig1}(a). We show that, despite the absence of interactions in the coherent dynamics, resetting induces quantum correlations as well as a  critical (non-analytic) behavior in the NESS. We demonstrate this by envisaging three distinct protocols, named henceforth Protocol \RomanNumeralCaps{1}, \RomanNumeralCaps{2} and \RomanNumeralCaps{3}, in increasing order of complexity [see Fig.~\ref{fig:fig1}(b,c)]. Protocol \RomanNumeralCaps{1} amounts to the aforementioned simple stochastic resetting of the system to a fixed state, while Protocols \RomanNumeralCaps{2} and \RomanNumeralCaps{3} include a measurement step whose outcome determines to which state the system is reset.

In \emph{all three} cases we find that resetting induces  long-range correlations, although the system's reset-free dynamics is non-interacting. These correlations, emerging from the global operations associated with the reset events, are not exclusively of statistical nature but also have a quantum origin. Moreover, Protocols \RomanNumeralCaps{2} and \RomanNumeralCaps{3} induce stationary collective behavior, which manifests in non-analyticities in an appropriate order parameter. While reminiscent of a  non-equilibrium phase transition, the phenomenology we observe here is rather different in nature. Standard phase transitions take place between phases with short-range correlations and finite susceptibility parameter. Here, instead, due to the reset process, the system features strong long-range correlations and a divergent susceptibility throughout the whole phase diagram and not only at the critical point. The collectively enhanced response of the system to external parameter variations  may  be exploited for high-density quantum sensing, as discussed, e.g., in Ref.~\cite{quantum_sensing}. The fact that such property emerges even within a simple non-interacting system readily realizable with neutral atoms highlights a novel and simple way for creating and exploiting correlated many-body states on quantum simulators \cite{Georgescu_2014,endres_2016,barredo_2016,robens_2017,Henriet2020quantumcomputing}.

\textbf{Dynamics and reset states.---}
We consider a system of $N$ spins with Hamiltonian
\begin{equation}
H = \Omega \sum_{i=1}^N \sigma_i^x + \Delta \sum_{i=1}^N \sigma_i^z ,
\label{mt:Ham}
\end{equation}
describing, for instance, non-interacting atoms subject to an external laser field. Here, $\sigma_i^{x,y,z}$ are the Pauli matrices of the $i$-th spin, $\Omega$ is the Rabi frequency and $\Delta$ is the laser detuning. The two basis states of each spin, $\ket{\uparrow}$ and $\ket{\downarrow}$, are chosen as the eigenstates of $\sigma^z$ and represent the ground state and the excited state, respectively [see Fig.~\ref{fig:fig1}(a)]. These can be, for example, two hyper-fine levels of an atom or of an ion.

Before turning to the discussion of the reset protocols, it is useful to first characterize the dynamical properties of the system during its coherent evolution. Since Hamiltonian~\eqref{mt:Ham} is the sum of single-body terms, we can  focus on the time evolution of single-body operators. For example, the local excitation density at site $j$, defined as $n_j = (1+\sigma_j^z)/2$, evolves as $n_j^F(t) = e^{i H_j t} n_j e^{-i H_j t}$ with $H_j = \Omega \sigma_j^x + \Delta \sigma_j^z$ and $F$ indicating evolution under the Hamiltonian reset-free dynamics. Without loss of generality, we fix the initial state to be $\ket{\uparrow}^N = \otimes_{i=1}^N \ket{\uparrow}_i$. With this choice one finds $\braket{n_j^F (t)}_\uparrow = 1 - (\Omega^2/\overline{\Omega}^2) \sin^2(\overline{\Omega}t)$, where $\overline{\Omega} = \sqrt{\Omega^2+\Delta^2}$ is the effective Rabi frequency and the arrow in the subscript indicates the initial state. 

The reset protocols are depicted in Fig.~\ref{fig:fig1}(b,c). All have in common that the system evolves coherently with Hamiltonian~\eqref{mt:Ham} in between consecutive reset events. In Protocol \RomanNumeralCaps{1}, we employ stochastic resetting, i.e., the system is reinitialized to the state $\ket{\uparrow}^N$ unconditionally to any measurement. In Protocols \RomanNumeralCaps{2} and \RomanNumeralCaps{3}, instead, the reset state is chosen conditionally on a measurement taken right before resetting, as pictured in Fig.~\ref{fig:fig1}(c). A natural choice for the quantity to be measured is the excitation density $n = \left(1/N \right)\sum_{i=1}^N n_i$. In particular, in Protocol \RomanNumeralCaps{2}, first proposed in Ref.~\cite{Gabriele_reset}, two reset states are present, $\ket{\uparrow}^N$ and $\ket{\downarrow}^N$, which correspond to the two completely polarized states with excitation density $1$ and $0$, respectively. The outcome of the measurement determines the reset state: if the measured excitation density exceeds a certain threshold, which is fixed to be $1/2$, then the system is reset to $\ket{\uparrow}^N$, otherwise it is reset to $\ket{\downarrow}^N$. In Protocol \RomanNumeralCaps{3}, the system is reset to $\ket{\uparrow}^N$ if the measured density exceeds the threshold. Otherwise, the coherent dynamics resumes from the state obtained by flipping all the spins in the post-measurement configuration, as sketched in Fig.~\ref{fig:fig1}(c). 

\textbf{Protocol \RomanNumeralCaps{1}: Unconditional reset.---}
In this case the coherent dynamics of the system is interrupted at random times at which the system is reset to  state $\ket{\uparrow}^N$. Resets happen at a constant rate $\gamma$. The time $\tau$ between consecutive reset is therefore distributed according to the Poisson waiting time distribution $f(\tau) = \gamma e^{-\gamma \tau}$. In principle this choice allows for arbitrarily long waiting times. In practice, i.e., in an experiment, a more appropriate waiting time distribution is the ``chopped exponential'' with a fixed maximum reset time \cite{Gabriele_reset,thermodynamics_reset}. This case is briefly discussed in the Supplemental Material~\footnote{See Supplemental Material for details of the calculations.} since it does not bear any additional conceptual difficulty with respect to the Poisson distribution. 

In the latter case, the survival probability, i.e., the probability that no reset happens for a time $\tau$, is given by $q(\tau) = \int_\tau^\infty f(s) ds = e^{-\gamma \tau}$. This, together with the reset-free time-evolved density matrix $\rho^F_\uparrow(t)$, determines the quantum state of the system $\rho_\uparrow(t)$ in the presence of resetting through the {\it last renewal} equation derived in Ref.~\cite{Mukherjee_2018}:
\begin{equation}
\rho_\uparrow(t) = e^{-\gamma t} \rho^F_\uparrow(t) + \gamma \int_0^t dt^\prime e^{-\gamma t^\prime} \rho^F_\uparrow(t^\prime).
\label{mt:master_eq}
\end{equation}
The first term in the above equation corresponds to having no reset up to time $t$. The second term accounts for realizations of the stochastic resetting process where the last reset has been at a previous time $t-t^\prime$ and the system has then evolved without reset events up to time $t$ via the Hamiltonian~\eqref{mt:Ham}.

\begin{figure*}
    \centering
    \includegraphics[width=0.9\linewidth]{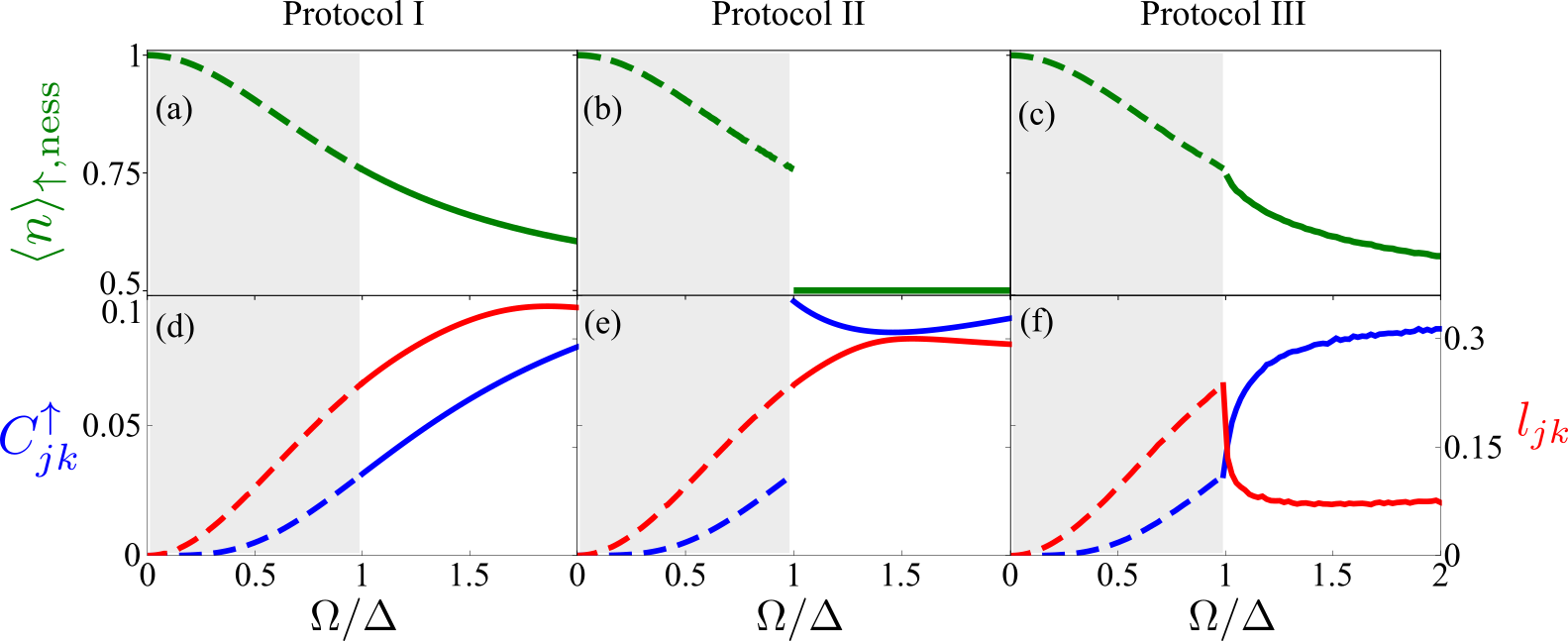}
    \caption{\textbf{Collective behavior and quantum correlations induced by reset.} First row: stationary excitation density as a function of $\Omega/\Delta$ for the three protocols. (a) For Protocol \RomanNumeralCaps{1} the order parameter (excitation density) is given by Eq.~\eqref{mt:n_stat_one_reset}. For Protocols \RomanNumeralCaps{2} (b) and \RomanNumeralCaps{3} (c), the order parameter displays a non-analyticity  at the critical point $\Omega_c = \Delta$, which is discontinuous or continuous, respectively. For protocol \RomanNumeralCaps{3} the order parameter behaves as a power law when approaching the critical point from the right  with an exponent close to $0.5$. Second row: connected correlation function (in blue, left axis) and quantum discord (in red, right axis), computed from the two-spin reduced density matrix $\rho_{jk}$, as a function of $\Omega/\Delta$. In contrast to Protocol \RomanNumeralCaps{1} (d), where both quantities are continuous, Protocol \RomanNumeralCaps{2} (e), leads to a discontinuity of both quantities at the critical point $\Omega_c = \Delta$. Note, that the discontinuity of the quantum discord is imperceptible on the scale shown. (f) For reset Protocol \RomanNumeralCaps{3}, both the connected correlation function and the quantum discord  feature power-law behavior in a right neighborhood of the critical point. The characteristic exponent is approximately $0.5$ for the connected correlation function and $0.2$ for the quantum discord. The dashed parts of the curves in all panels highlight the fact that, when $\Omega<\Delta$, the three protocols become equivalent. All data are obtained analytically, except for panels (c) and (f) where numerical simulations are necessary. The reset rate is chosen to be $\gamma = \Delta/2$.}
    \label{fig:fig2}
\end{figure*}

The average excitation density in the state (\ref{mt:master_eq}) is given by $\braket{n(t)}_\uparrow = \Tr[n \rho_\uparrow(t)]$ and its stationary value reads
\begin{equation}
\braket{n}_{\uparrow,\mathrm{ness}} = \lim_{t \rightarrow \infty} \braket{n(t)}_\uparrow = 1 - 2 \frac{\Omega^2}{\gamma^2 + 4 \overline{\Omega}^2},
\label{mt:n_stat_one_reset}
\end{equation}
which is shown in Fig.~\ref{fig:fig2}(a). This expression smoothly varies with $\Omega/\Delta$ contrary to what we will show for Protocols \RomanNumeralCaps{2} and \RomanNumeralCaps{3}. Equation~\eqref{mt:n_stat_one_reset} 
is equal to 1, i.e., the excitation density of the initial state, for $\Omega = 0$ (no coupling between single spin states), $\gamma \rightarrow \infty$ (the infinitely frequent resets induce a quantum Zeno effect \cite{zeno_effect_1,zeno_effect_2} which freezes the system to its initial state) and $\Delta \rightarrow \infty$ (transitions between the two spins states are highly off-resonant). 

Rather surprisingly, although in each realization of the process the system is in a product state at all times, the reset mechanism introduces long-range correlations. This becomes evident when looking at the stationary two-spin connected correlation function $C_{jk}^{\uparrow} = \left[ \braket{n_j n_k}_{\uparrow,\mathrm{ness}} - \braket{n_j}_{\uparrow,\mathrm{ness}} \braket{n_k}_{\uparrow,\mathrm{ness}} \right]$,  which is equal to
\begin{equation}
C_{jk}^{\uparrow} = 4 \Omega^4\frac{ 5 \gamma^2 + 8 \overline{\Omega}^2}{\left(\gamma^2 + 4\overline{\Omega}^2 \right)^2 \left(\gamma^2 + 16\overline{\Omega}^2 \right)} , 
\label{mt:corr_prot1}
\end{equation}
showing that correlations do not depend on  the considered spins. This is reminiscent of what happens in fully connected models (see, e.g., \cite{benatti2018} for an example in dissipative settings). However, in our case, these correlations are {\it strong} in the sense that they do not vanish in the thermodynamic $N\to\infty$ limit. As such, contrary to the case of fully connected models \cite{benatti2016}, the stationary state of our reset process is not clustering, i.e., it does not possess Gaussian fluctuations, as shown by the fact that the susceptibility is diverging: $\chi=\lim_{N\to\infty}1/N\sum_{j,k=1}^N C_{jk}^{\uparrow}=\infty$. 

In addition to these strong classical density-density correlations, the NESS, in fact, also contains correlations of quantum origin. This aspect can be shown by computing the local quantum uncertainty (LQU), defined in Ref.~\cite{LQU_Adesso}, which is a type of bipartite quantum discord \cite{quantum_discord_1,quantum_discord_2}. It quantifies the extent of the fluctuations of a local measurement due to the non-commutativity between the state and the measured local observable. The LQU isolates the fluctuations that are caused only by the coherence of the state and not by its mixedness. Here we compute the LQU for the stationary two-spin reduced density matrix $\rho_{jk}$ \cite{Note1}. It is given by $l_{jk} = 1 - \lambda_\mathrm{max}\{W_{jk}\}$, where $\lambda_\mathrm{max}\{W_{jk}\}$ is the largest eigenvalue of the $3 \times 3$ matrix $W_{jk}$ with elements $\left(W_{jk} \right)_{ab} = \Tr[\sqrt{\rho_{jk}} (\sigma_j^a \otimes \mathbb{1})\sqrt{\rho_{jk}} (\sigma_j^b \otimes \mathbb{1})]$, with $a,b = x,y,z$. As for the classical correlations, also the LQU does not depend on the distance between sites. In Fig.~\ref{fig:fig2}(d) we show the connected correlation function~\eqref{mt:corr_prot1} (left axis) together with the quantum discord quantified via the LQU (right axis) for Protocol \RomanNumeralCaps{1}. Both quantities possess qualitatively the same shape and smoothly vary with $\Omega/\Delta$.

\textbf{Protocol \RomanNumeralCaps{2}: Conditional reset to two states.---}
This protocol exploits two reset states: $\ket{\uparrow}^N$ and $\ket{\downarrow}^N$. At each reset event, the local density at each site is measured and the total excitation density $n$ is computed. The system is reinitialized to the reset state $\ket{\uparrow}^N$ if the majority of the spins is found in the excited state, i.e., $n>1/2$. On the contrary, if $n<1/2$, the  reset state is chosen as $\ket{\downarrow}^N$. For large $N$, the probability distribution for measuring a certain value of $n$ after a time $t$ since the last reset is a Gaussian distribution centered on the average $\braket{n^F(t)}_{\uparrow/\downarrow}$, with variance $\sigma_n^2 \propto 1/N$ \cite{Note1}. This means that, at each reset event, the system can in principle be reinitialized in both reset states, albeit with different probabilities. This aspect, together with the fact that the Hamiltonian dynamics of the average density satisfies the relation $\braket{n^F(t)}_\uparrow = 1 - \braket{n^F(t)}_\downarrow$, makes the stationary excitation density exactly equal to $1/2$, i.e., the average between the density of the two reset states \cite{Note1}.

A different phenomenology takes place in the thermodynamic limit $N \rightarrow \infty$. In this case, as a consequence of the \emph{law of large numbers} applied to the operator $n$, the probability distribution to measure a certain value for $n$ becomes a delta-function peaked around the average $\braket{n^F(t)}_{\uparrow/\downarrow}$. This self-averaging property makes the measurement of the excitation density fully deterministic with outcome equal to its average value. As a consequence, for $\Omega < \Delta$, given the initial condition and the fact that  $\braket{n^F(t)}_\uparrow > 1/2$ $ \; \forall t$, the system can only be reset to the state $\ket{\uparrow}^N$ and, therefore, the average  density in the process is always larger than $1/2$. For $\Omega > \Delta$, instead, both reset states can be reached so that  the stationary excitation density is equal to $1/2$ \cite{Note1}. The stationary excitation density, acting as an order parameter, then displays a jump discontinuity at the critical point $\Omega_c = \Delta$, as shown in Fig.~\ref{fig:fig2}(b). This is a consequence of an abrupt change in the dynamics: for $\Omega > \Delta$ the system can reset to both states, while for $\Omega < \Delta$ the dynamics is effectively that of Protocol \RomanNumeralCaps{1}, with the stationary excitation density coinciding with Eq.~\eqref{mt:n_stat_one_reset} [see also Fig.~\ref{fig:fig2}]. 

As shown in Fig.~\ref{fig:fig2}(e) the connected correlation function and the quantum discord display a behavior that is qualitatively different to that of Protocol \RomanNumeralCaps{1}. They are both discontinuous at the critical point even though the discontinuity of the LQU  is tiny on the scale of the figure.

\textbf{Protocol \RomanNumeralCaps{3}: Conditional reset to the initial state.---}
In the third protocol,  the system is reset to its initial state $\ket{\uparrow}^N$ only if the measured  excitation density exceeds $1/2$. If not, the system resumes its dynamics from the state generated by the projective measurement  after a subsequent flip of all its spins is performed [see Fig.~\ref{fig:fig1}(b,c)]. This means that, if the state after the projective measurement possesses an excitation density equal to $n^\prime < 1/2$, the reset state will have excitation density $1 - n^\prime > 1/2$. This protocol is still conditioned on the measured excitation density, but, in contrast to Protocol \RomanNumeralCaps{2}, any state with $n > 1/2$ can be considered as a reset state according to the parameter regime. The resulting non-equilibrium phase diagram, see  Fig.~\ref{fig:fig2}(c), exhibits a continuous non-analytic behavior at the critical point $\Omega_c = \Delta$. 

We note that, without the additional spin-flip operation, the stationary behavior of the density would be discontinuous also for this protocol. Indeed, when $\Omega > \Delta$, each realization  of the reset process would spend on average half of the time in configurations with $n$ smaller than $1/2$ and half of the time in configurations with $n$ larger than $1/2$. The stationary state, obtained by averaging over trajectories, would  therefore be  very different from the one attained when $\Omega < \Delta$, where trajectories maintain a positive magnetization, $n>1/2$, throughout the whole reset process. This substantial dissimilarity between the two regimes would result in a jump discontinuity of the order parameter at $\Omega_c$. On the contrary, with the introduction of the spin-flip operation, the order parameter is continuous, but still non-analytic, since its first derivative has a jump discontinuity at $\Omega_c$.
This can be understood by noticing that, in this case, for $\Omega \gsim \Delta$ each trajectory of the reset process spends only an infinitesimal time in states with $n < 1/2$, since after a reset the system restarts the dynamics from a state with $n > 1/2$. 

In the vicinity of $\Omega_c$, the order parameter displays a power-law behaviour $ \sim(\Omega - \Omega_c)^\beta$, for $\Omega \rightarrow \Omega_c^+$, with a static exponent $\beta \approx 0.5$. This  seems to indicate the emergence of a second-order phase transition in the NESS. However, looking at the behavior of the correlation function reveals a rather unexpected phenomenology. Indeed, in  second-order phase transitions, upon approaching the critical point, the correlation length of the system increases giving rise to a power-law divergence of the susceptibility at criticality. Here, instead, as mentioned already when discussing Protocol \RomanNumeralCaps{1}, the system features strong long-range correlations which  determine a divergence of the susceptibility parameter $\chi$ for any value of $\Omega/\Delta$ and not only at criticality. Despite this divergence, we can still analyse the two-spin correlation function $C_{jk}^{ \uparrow}$. This quantity, displayed in Fig.~\ref{fig:fig2}(f), interestingly also obeys a power-law behaviour $\sim (\Omega - \Omega_c)^\beta$ close to the critical point, with the same static exponent $\beta$ of the order parameter. Also the quantum discord, as measured by the LQU, follows a power law with exponent $\delta \approx 0.2$.

\textbf{Conclusions and outlook.---} We have shown that combining a non-interacting quantum dynamics with an externally imposed reset process can lead to surprisingly rich non-equilibrium stationary states. Even the simplest possible protocol results in a state with non-trivial classical and quantum correlations. More involved protocols lead to the emergence of a phase-transition behavior in an initially non-interacting system, which may be relevant for the implementation of quantum sensing and metrology applications~\cite{Wade2018,quantum_sensing,Jau_2020,Downes_2020}. The required non-linearities for such collective behavior are of course  generated by the reset which operates non-locally on the system.
Conceptually this may appear simpler than the creation of strong coherent interactions, which typically underlie collective dynamics. However, one requires the ability to rapidly read out and initialize the spin ensemble. For the results discussed in Fig.~\ref{fig:fig2} we have assumed a reset rate $\gamma = \Delta/2$, which in some settings may be impractical (it could be on the order of MHz for cold atoms). However, our findings do not change qualitatively for smaller values of the reset rate. The key quantity is indeed the ratio $\Omega/\Delta$, while the value of $\gamma$ simply provides the timescale for the approach to stationarity. 

\acknowledgments \textbf{Acknowledgements.---} G.P. acknowledges support from the Alexander von Humboldt Foundation through a Humboldt research fellowship for postdoctoral researchers. We acknowledge support from the “Wissenschaftler R\"uckkehrprogramm GSO/CZS” of the Carl-Zeiss-Stiftung and the German  Scholars Organization e.V., from the European Union's Horizon 2020 research and innovation program under Grant Agreement No.~800942 (ErBeStA), as well as from the Baden-W\"urttemberg Stiftung through Project No.~BWST\_ISF2019-23. We also acknowledge funding from the Deutsche Forschungsgemeinschaft through SPP 1929 (GiRyd), Grant No. 428276754.

\bibliography{bib}

\setcounter{equation}{0}
\setcounter{figure}{0}
\setcounter{table}{0}
\makeatletter
\renewcommand{\theequation}{S\arabic{equation}}
\renewcommand{\thefigure}{S\arabic{figure}}
\renewcommand{\bibnumfmt}[1]{[S#1]}
\renewcommand{\citenumfont}[1]{S#1}

\onecolumngrid
\newpage

\begin{center}
{\Large SUPPLEMENTAL MATERIAL}
\end{center}
\begin{center}
\vspace{0.8cm}
{\Large Emergent quantum correlations and collective behavior  in non-interacting quantum systems subject to stochastic resetting}
\end{center}

\begin{center}
Matteo Magoni$^{1}$, Federico Carollo$^{1}$, Gabriele Perfetto$^{1}$, and Igor Lesanovsky$^{1,2}$
\end{center}
\begin{center}
$^1${\em Institut für Theoretische Physik, Eberhard Karls Universität Tübingen,}\\
{\em Auf der Morgenstelle 14, 72076 T\"ubingen, Germany}\\
$^2$ {\em School of Physics and Astronomy and Centre for the Mathematics}\\
{\em and Theoretical Physics of Quantum Non-Equilibrium Systems,}\\
{\em The University of Nottingham, Nottingham, NG7 2RD, United Kingdom}
\end{center}

{Here} we derive in detail the results presented in the main text. The Supplemental Material is organized as follows. In Section~\ref{sec:sec1}, we report the general expression for the stationary density matrix that is valid for Protocol \RomanNumeralCaps{2}. In Section~\ref{sec:sec2}, we derive the probability distribution to observe a certain value of $n$ at a given time between two consecutive resets and discuss the stationary properties of Protocol \RomanNumeralCaps{2} when it is adopted in a finite system composed of $N$ spins. Sections~\ref{sec:sec3} and~\ref{sec:sec4} provide details on the computation of the quantum discord and the connected correlation function, respectively. Section~\ref{sec:sec5} briefly treats the case of non-Poissonian resetting, characterized by a fixed maximum reset time.

\section{General expression of the stationary density matrix for Protocol  \texorpdfstring{\RomanNumeralCaps{2}}{2}}
\label{sec:sec1}

The resetting dynamics described in Protocols \RomanNumeralCaps{1} and \RomanNumeralCaps{2} allows to write the exact form of the stationary density matrix $\rho_{\mathrm{ness}}$ in terms of the waiting time distribution and the reset-free dynamical properties of the system. In particular, for Protocol \RomanNumeralCaps{2}, the expression of the stationary density matrix $\rho_{\mathrm{ness}}$ has been determined in Ref.~[\onlinecite{Gabriele_reset}] and it reads as
\begin{equation}
\rho_{\mathrm{ness}} =   \frac{c_\uparrow}{\widehat{q}}\int_0^\infty dt^\prime q(t^{\prime}) \rho^F_\uparrow (t^\prime) + \frac{c_\downarrow}{\widehat{q}}  \int_0^\infty dt^\prime q(t^{\prime}) \rho^F_\downarrow (t^\prime),
\label{sm:stat_density_matrix}
\end{equation}
where
\begin{equation}
c_\uparrow = \frac{R_{\downarrow \uparrow}}{R_{\downarrow \uparrow}+R_{\uparrow \downarrow}}, \qquad \qquad \qquad
c_\downarrow = \frac{R_{\uparrow \downarrow}}{R_{\downarrow \uparrow}+R_{\uparrow \downarrow}},
\label{sm:weights}
\end{equation}
and
\begin{equation}
\widehat{q}=\int_{0}^{\infty} dt^{\prime} q(t^{\prime}),    \qquad \qquad \qquad R_{ij} = \gamma \int_0^\infty dt^\prime e^{-\gamma t^\prime} P_{ij}(t^\prime), \qquad \qquad i,j=\uparrow,\downarrow.
\label{sm:R_ij}
\end{equation}
In the previous equation, $P_{ij}(t)$ is the probability that the system, starting its reset-free evolution from the reset state $\ket{i}$ ($\ket{\uparrow}^N$ or $\ket{\downarrow}^N$), in the occurrence of a reset event after a time $t$, is reinitialized to the reset state $\ket{j}$ ($\ket{\uparrow}^N$ or $\ket{\downarrow}^N$). Equation~\eqref{sm:stat_density_matrix} expresses $\rho_{\mathrm{ness}}$ as a statistical mixture of the unitary time evolutions ensuing from the reset states $\ket{\uparrow}^N$ and $\ket{\downarrow}^N$. Fundamentally, both weights $c_\uparrow$ and $c_\downarrow$ couple the Hamiltonian dynamics with the reset via Eqs.~\eqref{sm:weights} and~\eqref{sm:R_ij}. In particular, since the probabilities $P_{ij}(t)$ depend on $\Omega$, also the weights $c_{\uparrow/\downarrow}$ depend on $\Omega$. 

In the main text and further below in Secs.~\ref{sec:sec2}-\ref{sec:sec4}, for the sake of simplicity, we consider the case of Poissonian resetting, with survival probability $q(t)=\mbox{exp}(-\gamma t)$, while we comment in Sec.~\ref{sec:sec5} about the non-Poissonian case. 

For Protocol \RomanNumeralCaps{3}, any state with excitation density $n>1/2$ can be considered as a reset state. The generalization of Eq.~\eqref{sm:stat_density_matrix} is therefore of no practical utility since it involves a summation over all the reset states, whose number is exponentially large in the system size. In order to obtain the stationary values of different properties such as the excitation density, the two-point correlation function and the quantum discord, in Protocol \RomanNumeralCaps{3}, we shall therefore resort to Monte Carlo simulations and use combinatorial properties (see Section \ref{sec:sec4}).

We finally note that the expression in Eq.~\eqref{sm:stat_density_matrix} does not apply  in the regime $\Omega<\Delta$, when considering the non-interacting spin system in the thermodynamic limit $N \to \infty$. Indeed, as we discuss below, for $\Omega<\Delta$, the magnetization of the spin ensemble can never change sign so that whether $n<1/2$ or $n>1/2$ throughout the whole dynamics solely depends on the value of $n$ in the initial state. This implies that the system cannot visit all the reset states, as witnessed for instance by the fact that $P_{\uparrow\downarrow}=P_{\downarrow\uparrow}=0$ for Protocol \RomanNumeralCaps{2} (similar relations would apply to Protocol \RomanNumeralCaps{3}). As such, the quantities $c_{\uparrow/\downarrow}$ become in principle ill-defined. In any case, it is straightforward to see that, starting from the state with all spins pointing up, in the regime $\Omega<\Delta$ and in the thermodynamic limit $N\to\infty$, the system can only reset to its initial state. As such, in this regime the stationary density matrix is the one given in Eq.~\eqref{sm:stat_density_matrix} with $c_{\uparrow}=1$ and $c_{\downarrow}=0$. Note that, in this limit, the stationary density matrix in Eq.~\eqref{sm:stat_density_matrix} reduces to the stationary limit of Eq.~(2) in the main text, as expected. This applies to both Protocol \RomanNumeralCaps{2} and Protocol \RomanNumeralCaps{3}.

\section{Statistical properties of the excitation density in a finite system}
\label{sec:sec2}

From Sec.~\ref{sec:sec1} it is evident that, once the probabilities $P_{ij}(t)$ in Eq.~\eqref{sm:R_ij} are computed, one can then easily obtain the stationary density matrix defined in Eq.~\eqref{sm:stat_density_matrix}, which is valid for Protocol \RomanNumeralCaps{2}. Exploiting the fact that the spins do not interact, it is indeed possible to compute those probabilities. Let us therefore focus on Protocol \RomanNumeralCaps{2}, where the measurement of the excitation density $n$ determines the reset state to choose. In particular, if the outcome of the measurement exceeds the threshold $1/2$, the selected reset state is $\ket{\uparrow}^N$, otherwise it is $\ket{\downarrow}^N$. It would be therefore beneficial to have an expression for the probability to measure a certain value of $n$ which, at a given time $t$, is above or below this threshold. To compute this probability it is of course sufficient to consider only the properties of the reset-free dynamics. The fact that the system is non interacting reduces the computation to a simple combinatorial problem. Since the threshold is $1/2$, a sort of majority rule applies in the sense that the threshold is exceeded whenever there are more up spins than down spins.  

As an example, let us show how to compute $P^{(N)}_{\uparrow \downarrow}(t)$, which is defined to be the probability that the system, being initialized in $\ket{\uparrow}^N$ (appearing as first subscript), is found after a time $t$ to have an excitation density $n<1/2$ (appearing as second subscript). In the notation of Sec.~\ref{sec:sec1}, it would be $P_{ij}(t)$, with $i=\uparrow$ and $j=\downarrow$. Following the majority rule, this amounts to the probability of having after a time $t$ more down spins than up spins. Assuming for simplicity that the total number of spins $N$ is odd and denoting with $p^{\uparrow \downarrow}(t) = (\Omega^2/\overline{\Omega}^2) \sin^2(\overline{\Omega}t)$ the probability that a single spin, initialized in the state $\ket{\uparrow}$ is found after a time $t$ in the state $\ket{\downarrow}$,
\begin{equation}
P^{(N)}_{\uparrow \downarrow}(t) = \sum_{k=0}^{\frac{N-1}{2}} {N \choose k} [1-p^{\uparrow \downarrow}(t)]^k [p^{\uparrow \downarrow}(t)]^{N-k},
\label{sm:binomial}
\end{equation}
which takes into account all the possible spin configurations with at most $(N-1)/2$ spins in the excited state. By using the normal approximation of the binomial distribution, which is valid for large $N$, see, e.g., Ref.~[\onlinecite{feller1957introduction}],
\begin{equation}
{N \choose k} (1-p)^k p^{N-k} \simeq \frac{1}{\sqrt{2 \pi N p (1-p)}} \exp{\left[-\frac{[k-N(1-p)]^2}{2Np(1-p)}\right]},
\end{equation}
and approximating the discrete sum with an integral, Eq.~\eqref{sm:binomial} gets simplified to
\begin{equation}
P^{(N)}_{\uparrow \downarrow}(t) \simeq \frac{1}{\sqrt{2 \pi N p^{\uparrow \downarrow}(t) [1-p^{\uparrow \downarrow}(t)]}} \int_0^\frac{N}{2} dx \exp\left\{-\frac{[x-N(1-p^{\uparrow \downarrow}(t))]^2}{2Np^{\uparrow\downarrow}(t)[1-p^{\uparrow \downarrow}(t)]}\right\}.
\end{equation}
We can therefore write the probability $P^{(N)}_{\uparrow \downarrow}(t)$ as a difference between two error functions as
\begin{equation}
P^{(N)}_{\uparrow \downarrow}(t) = \frac{1}{2} \left[\erf{\left(\frac{-\frac{N}{2} + N p^{\uparrow \downarrow}(t)}{\sqrt{2 N p^{\uparrow \downarrow}(t) [1-p^{\uparrow \downarrow}(t)]}}\right)} - \erf{\left(\frac{-N + N p^{\uparrow \downarrow}(t)}{\sqrt{2 N p^{\uparrow \downarrow}(t) [1-p^{\uparrow \downarrow}(t)]}}\right)} \right].
\label{sm:P_up_down}
\end{equation}
With the explicit expression for the probabilities $P_{ij}(t)$, of which Eq.~\eqref{sm:P_up_down} is an example, one can obtain the stationary density matrix~\eqref{sm:stat_density_matrix}. Note that, thanks to the fact that the single spin transition probabilities satisfy $p^{\uparrow \downarrow}(t) = p^{\downarrow \uparrow}(t)$,
\begin{equation}
P^{(N)}_{\uparrow \downarrow}(t) = P^{(N)}_{\downarrow \uparrow}(t) \qquad \text{and} \qquad P^{(N)}_{\uparrow \uparrow}(t) = P^{(N)}_{\downarrow \downarrow}(t),
\label{sm:symmetry_prob}
\end{equation}
so the symmetry in the reset-free dynamics is not restricted to the average value of the excitation density operator through $\braket{n^F(t)}_\uparrow = 1 - \braket{n^F(t)}_\downarrow$, but it is also extended to the probabilities.

In the thermodynamic limit, Eq.~\eqref{sm:P_up_down} can be further simplified. Indeed, for $N \rightarrow \infty$, the second term tends to 1 because $p^{\uparrow \downarrow}(t) \leq 1$. On the other hand, the first term tends to $+1$ if $p^{\uparrow \downarrow}(t) > 1/2$ or to $-1$ if $p^{\uparrow \downarrow}(t) < 1/2$. Note that the number $1/2$ comes from the chosen threshold. As a consequence, the probability $P^{(N)}_{\uparrow \downarrow}(t)$ simply reduces to a Heaviside step function with a time dependent argument
\begin{equation}
P^{(\infty)}_{\uparrow \downarrow}(t) = \lim_{N \rightarrow \infty} P^{(N)}_{\uparrow \downarrow}(t) = \Theta \left(p^{\uparrow \downarrow}(t) - \frac{1}{2} \right).
\label{sm:theta_func}
\end{equation}
Importantly, this result shows that $P^{(\infty)}_{\uparrow \downarrow}(t)$ can be either 1 or 0, meaning that, in the thermodynamic limit, the excitation density $n = (1/N)\sum_{i=1}^N n_i$, when measured, takes deterministically a certain value, which turns out to be equal to the average value of the single spin excitation density. This self-averaging property shows indeed that the fluctuations of $n$ around its average value are suppressed, in accordance with the \emph{law of large numbers}.

It is also interesting to see how the previous results change if $N$ is assumed to be large but finite. In particular, given the large $x$ expansion of the error function as $\erf{x} \simeq 1 - e^{-x^2}/(\sqrt{\pi} x)$, Eq.~\eqref{sm:theta_func} gets modified by a correction of order $e^{-N}/\sqrt{N}$ as
\begin{equation}
P^{(N)}_{\uparrow \downarrow}(t) \simeq \frac{\sqrt{2p^{\uparrow \downarrow}(t)[1-p^{\uparrow \downarrow}(t)]}}{2 \sqrt{\pi N}} \left(\frac{e^{-N\frac{\left[\frac{1}{2} - p^{\uparrow \downarrow}(t) \right]^2}{2 p^{\uparrow \downarrow}(t) [1-p^{\uparrow \downarrow}(t)]}}}{\frac{1}{2} - p^{\uparrow \downarrow}(t)} - \frac{e^{-N\frac{\left[1-p^{\uparrow \downarrow}(t) \right]^2}{2 p^{\uparrow \downarrow}(t)[1-p^{\uparrow \downarrow}(t)]}}}{1 - p^{\uparrow \downarrow}(t)} \right), \qquad \text{if} \; \; p^{\uparrow \downarrow}(t) < \frac{1}{2},
\label{sm:P_large_N_1}
\end{equation}
and 
\begin{equation}
P^{(N)}_{\uparrow \downarrow}(t) \simeq 1 - \frac{\sqrt{2p^{\uparrow \downarrow}(t)[1-p^{\uparrow \downarrow}(t)]}}{2 \sqrt{\pi N}} \left(\frac{e^{-N\frac{\left[\frac{1}{2} - p^{\uparrow \downarrow}(t) \right]^2}{2p^{\uparrow \downarrow}(t)[1-p^{\uparrow \downarrow}(t)]}}}{\frac{1}{2} - p^{\uparrow \downarrow}(t)} + \frac{e^{-N\frac{\left[1-p^{\uparrow \downarrow}(t) \right]^2}{2p^{\uparrow \downarrow}(t)[1-p^{\uparrow \downarrow}(t)]}}}{1 - p^{\uparrow \downarrow}(t)} \right), \qquad \text{if} \; \; p^{\uparrow \downarrow}(t) > \frac{1}{2}.
\label{sm:P_large_N_2}
\end{equation}
Note that for $N \rightarrow \infty$ one recovers the result in Eq.~\eqref{sm:theta_func}.

\begin{figure}
    \centering
    \includegraphics{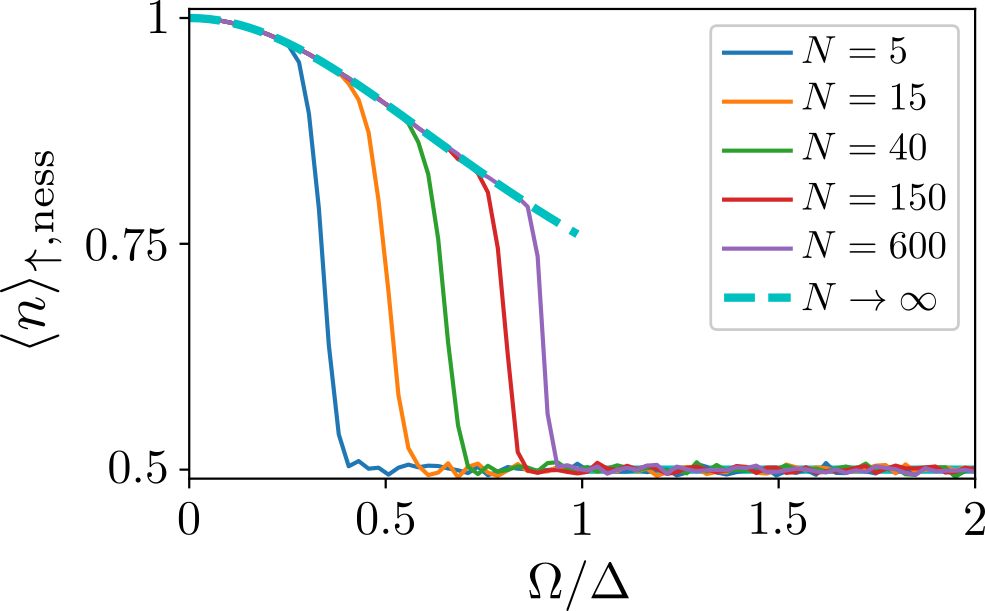}
    \caption{\textbf{Phase diagram of the (quasi)-stationary excitation density in Protocol \RomanNumeralCaps{2} in a finite system}. The first order phase transition that takes place in the thermodynamic limit becomes a crossover in a finite system. In this latter case, the stationary excitation density would be a constant function equal to $1/2$. Due to the exponentially small finite size corrections to Eq.~\eqref{sm:theta_func}, only a (quasi)-stationary state can be obtained numerically, which displays the expected plateau up to a certain value of $\Omega/\Delta$. The plots are obtained numerically averaging over 10000 trajectories of the reset process. The reset rate is $\gamma = \Delta/2$ and the observation time is $T=2000$ in units of $1/\Delta$. The dashed line, valid in the thermodynamic limit, is Eq.~(3) of the main text.}
    \label{sm:different_N}
\end{figure}

Figure~\ref{sm:different_N} investigates the behavior of the order parameter in Protocol \RomanNumeralCaps{2} as a function of $\Omega/\Delta$ for various numbers $N$ of particles.
The plotted curves are obtained with Monte Carlo simulations. In particular, we fix a large observation time $T$ and we simulate several realizations of the reset process within this time interval by drawing the times between consecutive resets from the waiting time distribution $f(\tau)$. The average of the computed excitation density at time $T$ over the many independent realizations of the reset process gives the numerical estimate of $\braket{n}_{\uparrow, \mathrm{ness}}$. This procedure is repeated for different values of $\Omega/\Delta$ leading to the result in Fig.~\ref{sm:different_N}. The discontinuous non-analytic behavior of the order parameter occurring in Protocol \RomanNumeralCaps{2}, and shown in Fig.~2(b) of the main text, becomes a continuous crossover when $N$ is finite. The reason for the observed smoothening is due to the fact that for finite $N$ the measurement of $n$ is no more deterministic and does not coincide with its average value because of the statistical fluctuations encoded in Eqs.~\eqref{sm:P_large_N_1} and \eqref{sm:P_large_N_2}. As a consequence, even for $\Omega < \Delta$, the probability to measure $n<1/2$ is nonzero and the system can be reset to the state $\ket{\downarrow}^N$. Because of the symmetry relation between transition probabilities given by Eq.~\eqref{sm:symmetry_prob}, both coefficients $c_\uparrow$ and $c_\downarrow$ in Eq.~\eqref{sm:stat_density_matrix} would be equal to $1/2$, leading to a stationary value of the excitation density, computed as $\Tr[n \rho_\mathrm{ness}]$, equal to $1/2$ for any value of $\Omega/\Delta$. This is only partly captured in Fig.~\ref{sm:different_N}, because the exponentially small correction \eqref{sm:P_large_N_1} and \eqref{sm:P_large_N_2} to Eq.~\eqref{sm:theta_func} due to finite size effects would require an exponentially long simulation to make this effect visible. In other words, obtaining numerically the stationary state for finite $N$ becomes challenging, because an exponentially large value of $T$ is needed. Nevertheless, for small values of $N$, the aforementioned correction becomes larger, making the predicted plateau more visible as the crossover tends to take place at smaller values of $\Omega/\Delta$. However, since these long timescales are hardly reached in current experiments due to dissipative and incoherent effects, the curves of Fig.~\ref{sm:different_N} resemble what can be realistically observed in the laboratory.

A change from the continuous non-analytic behavior  to a smooth crossover is also expected to happen in Protocol \RomanNumeralCaps{3}, although the stationary excitation density would not be equal to $1/2$, but would remain a decreasing function of $\Omega/\Delta$, because all the possible reset states possess a positive magnetization ($n>1/2$) and the reasoning that makes use of the symmetric relation~\eqref{sm:symmetry_prob} cannot be exploited.

\section{Computation of the quantum discord}
\label{sec:sec3}

In the main text we compute the quantum discord of the stationary two-spin reduced density matrix, defined as
\begin{equation}
\rho_{jk} = \lim_{t \rightarrow \infty} \rho_{jk}(t),
\end{equation}
where $\rho_{jk}(t)$ is the two-spin reduced density matrix at time $t$. In Protocol \RomanNumeralCaps{1} it is possible to compute explicitly $\rho_{jk}(t)$ which, from Eq.~(2) of the main text, is given by
\begin{equation}
\rho_{jk}(t) = e^{-\gamma t} \rho_{jk,\uparrow}^F(t) + \gamma \int_0^t dt^\prime e^{-\gamma t^\prime} \rho_{jk,\uparrow}^F(t^\prime),
\label{sm:rho_reduced_finite_t}
\end{equation}
where $\rho_{jk,\uparrow}^F(t) = \rho_{j,\uparrow}^F(t) \otimes \rho_{k,\uparrow}^F(t) = (e^{-i H_j t} \ket{\uparrow}_j \bra{\uparrow}_j e^{i H_j t}) \otimes (e^{-i H_k t} \ket{\uparrow}_k \bra{\uparrow}_k e^{i H_k t})$, with $H_j = \Omega \sigma_j^x + \Delta \sigma_j^z$.
The stationary reduced density matrix is therefore obtained by taking the infinite time limit
\begin{equation}
\rho_{jk} = \gamma \int_0^\infty dt^\prime e^{-\gamma t^\prime} \rho_{jk,\uparrow}^F(t^\prime),
\label{sm:stat_prot_1}
\end{equation}
which suppresses the first term of Eq.~\eqref{sm:rho_reduced_finite_t}. In Protocol \RomanNumeralCaps{2}, instead, $\rho_{jk}$ has two different expressions for $\Omega < \Delta$ and $\Omega > \Delta$. In particular, as also mentioned in the main text about the observable $n$, when $\Omega < \Delta$, its expression is the same of Eq.~\eqref{sm:stat_prot_1}, due to the equivalence between the two protocols. When $\Omega > \Delta$, instead, $\rho_{jk}$ takes its most general form from Eq.~\eqref{sm:stat_density_matrix} and is given by two contributions, referring to the reset-free evolution from the two reset states, weighted by the coefficients $c_\uparrow$ and $c_\downarrow$ as~[\onlinecite{Gabriele_reset}] 
\begin{equation}
\rho_{jk}= \gamma \left[c_\uparrow \int_0^\infty dt^\prime e^{-\gamma t^\prime} \rho_{jk,\uparrow}^F(t^\prime) + c_\downarrow \int_0^\infty dt^\prime e^{-\gamma t^\prime} \rho_{jk,\downarrow}^F(t^\prime) \right].
\label{sm:stat_prot_2}
\end{equation}
Because of the symmetric relation $\braket{n(t)}_\uparrow = 1 - \braket{n(t)}_\downarrow$, one has from Eq.~\eqref{sm:symmetry_prob} that $c_\uparrow$ and $c_\downarrow$ in Eq.~\eqref{sm:stat_density_matrix} simplify as $c_\uparrow = c_\downarrow = 1/2$. This implies that the two contributions are equally weighted. This is also the reason why, for $\Omega > \Delta$, $\braket{n}_{\uparrow, \mathrm{ness}} = 1/2$, as explained in Sec.~\ref{sec:sec2}. 

The reduced density matrices in Eqs.~\eqref{sm:stat_prot_1} and~\eqref{sm:stat_prot_2} are used to compute the quantum discord for Protocol \RomanNumeralCaps{1} (and Protocol \RomanNumeralCaps{2} for $\Omega < \Delta$) and Protocol \RomanNumeralCaps{2} (only for $\Omega > \Delta$), respectively. As mentioned in the main text, the quantum discord is quantified via the LQU, which is defined according to Ref.~[\onlinecite{LQU_Adesso}] as $l_{jk} = 1 - \lambda_\mathrm{max}\{W_{jk}\}$, where $\lambda_\mathrm{max}\{W_{jk}\}$ is the largest eigenvalue of the $3 \times 3$ matrix $W_{jk}$ with elements $\left(W_{jk} \right)_{ab} = \Tr[\sqrt{\rho_{jk}} (\sigma_j^a \otimes \mathbb{1})\sqrt{\rho_{jk}} (\sigma_j^b \otimes \mathbb{1})]$, with $a,b = x,y,z$.

\section{Computation of the connected correlation function}
\label{sec:sec4}

In the main text we also compute the connected correlation function between spins at site $j$ and $k$. Its stationary value is defined as
\begin{equation}
C_{jk}^{\uparrow} = \braket{n_j n_k}_{\uparrow,\mathrm{ness}} - \braket{n_j}_{\uparrow,\mathrm{ness}} \braket{n_k}_{\uparrow,\mathrm{ness}}.
\end{equation}
For Protocol \RomanNumeralCaps{1} and for Protocol \RomanNumeralCaps{2} when $\Omega < \Delta$, its value is given by Eq.~(4) of the main text. For Protocol \RomanNumeralCaps{2} when $\Omega > \Delta$, its expression can also be computed exactly using the stationary density matrix~\eqref{sm:stat_prot_2} and is given by
\begin{equation}
C_{jk}^{\uparrow} = \frac{1}{4} - 2 \Omega^2 \frac{\gamma^2 - 12 \Omega^2 + 16 \overline{\Omega}^2}{\gamma^4 + 20 \gamma^2 \overline{\Omega}^2 + 64 \overline{\Omega}^4}.
\label{sm:corr_func_2}
\end{equation}
This function is plotted in Fig. 2(e) of the main text.

As mentioned in Sec.~\ref{sec:sec1}, for Protocol \RomanNumeralCaps{3}, we resort to numerical Monte Carlo simulations to efficiently compute the connected correlation function and the quantum discord. We adopt the same numerical procedure described in Sec.~\ref{sec:sec2} by simulating 80000 independent realizations of the reset process up to the observation time $T = 30$ (in units of $1/\Delta$). The average over the reset realizations of the connected correlation function at time $T$ is plotted, as a function of $\Omega/\Delta$, in Fig.~2(f) of the main text. Note that, contrary to what has been previously observed for the estimate of $\braket{n}_{\uparrow, \mathrm{ness}}$, now there is no need for a very large value of $T$, because the simulations are done in the thermodynamic limit and, therefore, one does not need a long time to reach the stationary state.  What is needed is just the dynamics of $\braket{n_j^F(t) n_k^F(t)}$ between two consecutive resets. In Protocol \RomanNumeralCaps{3}, every state with positive magnetization ($n>1/2$) can be considered as a reset state. Therefore, we need an expression for the reset-free dynamics of the two-point correlation function for any possible initial state with $n_0 > 1/2$. The dynamics of the order parameter is readily obtained as 
\begin{equation}
\braket{n^F(t)}_{n_0} = n_0 \braket{n^F(t)}_\uparrow + (1-n_0) \braket{n^F(t)}_\downarrow,    
\end{equation}
because $N_0 = N n_0$ spins evolve starting from the $\ket{\uparrow}$ state and the remaining $N-N_0$ from the $\ket{\downarrow}$ state. For the two-point correlation function, the computation is slightly more complicated, because, given an initial state with excitation density $n_0$, both spins at sites $j$ and $k$ can be initialized to $\ket{\uparrow}$ or $\ket{\downarrow}$, so four different combinations are possible. Moreover, although the probability that a \textit{single} spin is initialized to $\ket{\uparrow}$ is exactly $n_0$, an analog reasoning cannot naively be applied for two spins, because the event that one spin is in the excited state is clearly not independent from the state of the other spin. As a consequence, one has to follow another procedure through direct counting. In particular, since the spin pair at sites $j$ and $k$ can be initialized in four possible ways, given an initial state with excitation density $n_0$, the dynamics of the two-point correlation function until the next reset event is given by
\begin{equation}
\braket{n_j^F(t) n_k^F(t)}_{n_0} = c_{\uparrow \uparrow} \braket{n_j(t)^F n_k(t)^F}_{\uparrow \uparrow} + c_{\uparrow \downarrow} \braket{n_j(t)^F n_k(t)^F}_{\uparrow \downarrow} + c_{\downarrow \uparrow} \braket{n_j(t)^F n_k(t)^F}_{\downarrow \uparrow} + c_{\downarrow \downarrow} \braket{n_j(t)^F n_k(t)^F}_{\downarrow \downarrow},
\label{sm:two_point}
\end{equation}
where the coefficients $c_{ab}$ are the probabilities to find the spin at site $j$ initialized in the state $\ket{a}$ and the spin at site $k$ initialized in the state $\ket{b}$, given that the system has excitation density $n_0$. In order to compute these probabilities, let us first count the number of possible spin configurations which give a total excitation density equal to $n_0 = N_0/N$. This is given by ${N \choose N_0}$. The number of configurations in which both spins at site $j$ and $k$ are in the excited state is obtained by counting the possible ways to arrange the remaining $N_0 - 2$ up spins among the remaining $N-2$ sites. Since this number is simply given by ${N-2 \choose N_0-2}$, the first coefficient entering Eq.~\eqref{sm:two_point} reads
\begin{equation}
c_\mathrm{\uparrow \uparrow} = \left. {N-2 \choose N_0-2} \middle/ {N \choose N_0} \right. = \frac{N_0 (N_0-1)}{N(N-1)}.
\end{equation}
Analogously, the other coefficients are given by
\begin{eqnarray}
c_\mathrm{\uparrow \downarrow} &=& \left. {N-2 \choose N_0-1} \middle/ {N \choose N_0} \right. = \frac{N_0 (N-N_0)}{N(N-1)}, \\ \nonumber
c_\mathrm{\downarrow \uparrow} &=& c_\mathrm{\uparrow \downarrow}, \\ \nonumber
c_\mathrm{\downarrow \downarrow} &=& \left. {N-2 \choose N_0}\middle/ {N \choose N_0} \right. = \frac{(N-N_0)(N-N_0-1)}{N(N-1)}.
\end{eqnarray}
One can check that the coefficients normalize to 1, i.e. $c_\mathrm{\uparrow \uparrow}+c_\mathrm{\uparrow \downarrow}+c_\mathrm{\downarrow \uparrow}+c_\mathrm{\downarrow \downarrow} = 1$. By taking the thermodynamic limit $N \rightarrow \infty$, their dependence on $N$ disappears and they reduce to
\begin{eqnarray}
c_\mathrm{\uparrow \uparrow} &=& n_0^2, \label{sm:coefficients_TDL}
\\ \nonumber
c_\mathrm{\uparrow \downarrow} &=& c_\mathrm{\downarrow \uparrow} = n_0 (1-n_0), \\ \nonumber
c_\mathrm{\downarrow \downarrow} &=& (1-n_0)^2,
\end{eqnarray}
showing that the thermodynamic limit eliminates the statistical dependence between the state of the spin at site $j$ and the one of the spin at site $k$. One can then easily obtain the dynamics of the correlation function by inserting these coefficients in Eq.~\eqref{sm:two_point} and exploiting the factorization $\braket{n_j^F(t) n_k^F(t)} = \braket{n_j^F(t)} \braket{n_k^F(t)}$ as a result of the fact that the spins do not interact. The quantum discord, plotted in Fig.~2(f) of the main text as a function of $\Omega/\Delta$, is obtained numerically in an analogous way by simulating 20000 independent realizations of the reset process up to the observation time $T = 30$ (in units of $1/\Delta$). Specifically, one needs the dynamics $\rho_{jk}^F(t)_{n_0}$ of the two-spin reduced density matrix between two consecutive resets. The dynamics $\rho_{jk}^F(t)_{n_0}$ is then written analogously as in Eq.~\eqref{sm:two_point} in terms of the reset-free dynamics $\rho_{jk}^F(t)_{ab}$, where the spins $j$ and $k$ are initialized in the state $\ket{a}$ and $\ket{b}$, respectively. The coefficients $c_{ab}$ of the four terms in the sum are again given in Eq.~\eqref{sm:coefficients_TDL}.

\section{Non-Poissonian resetting}
\label{sec:sec5}

In the main text and in the previous sections we focus on the Poissonian resetting, where the waiting time distribution is an exponential function. To account for the finite coherence time attained in cold-atom systems, a more suitable waiting time distribution would, however, be of the form of a ``chopped exponential''~[\onlinecite{Gabriele_reset}]
\begin{equation}
f(t) = \frac{\gamma}{1-e^{-\gamma t_\mathrm{max}}} e^{-\gamma t} \, \Theta(t_\mathrm{max}-t),
\label{sm:chopped_exp}
\end{equation}
where $t_\mathrm{max}$ is the maximum reset time. The survival probability then reads
\begin{equation}
q(t) = \frac{e^{-\gamma t}-e^{-\gamma t_\mathrm{max}}}{1-e^{-\gamma t_\mathrm{max}}} \, \Theta(t_\mathrm{max}-t).
\label{sm:chopped_surv}
\end{equation}
The non-Poissonian case of Eqs.~\eqref{sm:chopped_exp} and \eqref{sm:chopped_surv} does not bear any additional conceptual difficulty with respect to the Poissonian one and it can be analyzed along the same lines using Eq.~\eqref{sm:stat_density_matrix} (which is indeed valid for an arbitrary waiting time distribution $f(t)$ and survival probability $q(t)$). For Protocol \RomanNumeralCaps{1}, the stationary density matrix $\rho_{\mathrm{ness}}$ is obtained from the limiting form of Eq.~\eqref{sm:stat_density_matrix} with $c_{\uparrow}=1$ and $c_{\downarrow}=0$, as explained in Sec.~\ref{sec:sec1}.

The results remain qualitatively the same as in the Poissonian case, with the appearance of a discontinuous and a continuous non-analytic behavior of the order parameter at the same critical point $\Omega_c = \Delta$ in Protocol \RomanNumeralCaps{2} and \RomanNumeralCaps{3}, respectively. This is, in particular, true as long as $t_\mathrm{max}$ is large enough compared to $\Omega^{-1}$ to allow for the magnetization to change sign in the regime $\Omega>\Delta$. If this is not the case, then all the protocols reduce to Protocol \RomanNumeralCaps{1}.
The properties of the correlation function and the quantum discord also remain unchanged. As an example, we report here the expression of the stationary excitation density $\braket{n}_{\uparrow,\mathrm{ness}}$ for $\Omega < \Delta$ as
\begin{equation}
\braket{n}_{\uparrow,\mathrm{ness}} = 1 - \frac{\Omega^2}{2 \overline{\Omega}^2(\gamma^2+4 \overline{\Omega}^2)}\left\{4 \overline{\Omega}^2 - \frac{\gamma^2}{e^{\gamma t_\mathrm{max}}-1-\gamma t_\mathrm{max}} \left[2 \sin^2(\overline{\Omega} t_\mathrm{max})-\frac{\gamma}{\overline{\Omega}} \sin(\overline{\Omega} t_\mathrm{max}) \cos(\overline{\Omega} t_\mathrm{max}) + \gamma t_\mathrm{max} \right] \right\},
\end{equation}
which reduces to Eq.~(3) of the main text for $t_\mathrm{max} \rightarrow \infty$.

\end{document}